\documentclass[useAMS,usenatbib,referee]{mn2e}
\usepackage{graphicx}
\usepackage{times}

\title{A new representation of the light curve and its power density spectrum}

\author[J. Tian \& Y. C. Zou]{{Jun Tian, and Yuan-Chuan Zou}\\
{School of Physics, Huazhong University of Science and Technology, Wuhan 430074, China}
\thanks{Email: zouyc@hust.edu.cn (YCZ)}}

\begin{document}
\maketitle

\begin{abstract}
We present a new representation of light curves, which is quite different from the binning method. Instead of choosing uniform bins, the reciprocal of interval between two successive photons is adopted to represent
the counting rate. A primary application of this light curve is to compute the power density spectrum by means of Lomb Periodogram and to find possible periods. To verify this new representation, we apply this method to artificial periodic data and some known periodic celestial objects, and the periods are all correctly found. Compared with the traditional fast Fourier transform method, our method does not rely on the bin size and has a spontaneously high time resolution, guaranteeing a wide frequency range in power density spectrum, and is especially useful when the photons are rare for its little information losses.
Some other applications of the new light curve, like pulse identification, variability and spectral time lag, are also discussed.
\end{abstract}

\begin{keywords}
{gamma-rays: general --- methods: data analysis --- X-rays: general}
\end{keywords}

\section{Introduction}

In astronomy, time series analysis has always been a focus of attention.
% With more and more satellites \citep{KSO,BOE97,LAM04,MER03} sent to outer space, 
The application of automatic data acquisition systems emphasize the need to develop the data analysis method \citep{SCA81}. As an important kind of time series most frequently used, light curve provides us the most fundamental observation and a straightforward profile of various celestial objects evolving with time.

%After discovered by Vela in 1973 \citep{KSO}, GRB has been a controversial phenomenon, and the origin is still a puzzle. To explore this highly energetic transient, many more satellites \citep{BOE97,LAM04,MER03} are sent to outer space to get the most fundamental observation, light curve, which provides a straightforward profile of GRBs evolving with time.

%Within the prompt emission of GRBs, there are various degrees of variability with time scales ranging from milliseconds \citep{BHA92,WAL00} up to several seconds. In principle, an exact definition of the variability is helpful to constrain the activity of the central engine, as well as the radiation mechanism and dissipation processes, which is supposed to decide the prompt emission, light curve, and energy distribution, spectrum, of GRBs \citep{GHI11,ZHA11}. To investigate GRB temporal variability, Fourier transform is used to get the power density over certain frequencies \citep{BEL98}, i.e., variability over different time scales. The power density spectrum (PDS) contains lots of useful information, for example, periodic oscillation in the case of a detection of an excess in the PDS \citep{MAR09}, dominant frequency \citep{LAZ02} and maximum frequency of variability \citep{UKW11}. Despite of diverse slopes of the PDS \citep{BEL98}, an average slope of 5/3 is used to constrain the parameters in the context of the internal shock model \citep{REE94,PAN99,SPA00}.

In data processing of light curve, it is preferred to count the photons in a fixed time bin\citep[e.g.,][]{ELL06,EHU02}. However, this method ignores the time distribution of each individual photons in the bin and reduces the time resolution. Instead, if we use the reciprocal of intervals between two successive photons to denote the counting rate in this time to constitute a light curve,
%\footnote{Hereafter light curve without special declaration refers to the light curve so obtained}
time resolution will be enhanced dramatically in a natural way in the cost of uneven distribution in time domain, which is not so simple to handle as even case. Besides, with arrival time of every photon used, it may provide a different thread to show temporal behaviour of different astronomical phenomena and identify some subtle variable structure in the light curve. \citet{SCA08} has used the un-binned data to detect quantum gravity photon dispersion by means of calculating the Shannon entropy. It is found that the entirely bin-free data can avoid the information losses resulting from the predefined bins and is especially suitable for weak sources.

Within time profile of different celestial objects, there are various degrees of variability, e.g., in GRBs ranging from milliseconds \citep{BHA92,WAL00} up to several seconds. In principle, an exact definition of the variability is helpful to constrain the activity of the central bodies, as well as the radiation mechanism, dissipation processes and so on, which integrate together to decide the light curve (e.g., \citet{GHI11,ZHA11}). To investigate the temporal variability, Fourier transform is used to get the power density over certain frequencies \citep{BEL98}. The power density spectrum (PDS) contains lots of useful information, like periodic oscillation in the case of a detection of an excess in the PDS, dominant frequency \citep{LAZ02}, maximum frequency of variability \citep{UKW11} and the slope of the PDS \citep{BEL98}.

Temporal properties like periodic or quasi-periodic pulsations of various celestial bodies have been extensively studied with the time-binned light curve (e.g., \citet{MAZ79,VAN89,GUI12,DIC13}). As is well known, periodic oscillations are typical of the observed properties of pulsars and X-ray binaries. Even for GRBs, in despite of some uncertainties, a quasi-periodic oscillation with period 8.06s in GRB 090709A was also reported \citep{MAR09} (see however \citet{CEN10}). The periodic property has long been analysed in many astronomical observations, and is of great importance to interpret the activity of central objects and constrain models. Here we make use of the unevenly spaced light curve instead and calculate PDS by means of Lomb Periodogram \citep{LOM76,SCA82}. Compared with previous method, a significant improvement is a wider frequency range as a result of higher time resolution, which may reveal periodicity in high frequencies.

%Otherwise, no period was found in the PDS given by the fast Fourier transform (FFT) technique calculated in evenly spaced data \citep{BEL98,GUI12,DIC13}.

In this work, we define the counting rate of the light curve from the individual time tagged event (TTE) data, and then transform it from time to frequency domain by Lomb Periodogram to see the validity in searching the periodic signals. It is presented in section 2. In section 3, we apply this method to check the periodicity of artificial data and various celestial objects. In section 4, we analyse the data processing results and give our interpretation to their physical implications. Except for the PDS, some other applications of the new light curve representation are also discussed.

\section{Data Analysis Method}

%This section describes a new method for analysis of time-tagged data to draw a light curve and its PDS. We don't utilize the tagged energy information because energy influence on light curve and PDS is not considered here. And we will leave it as a subject in future works.

So far, most methods of data analysis require binning of the raw photon data in a set of predefined bins. Once the width of the bins is set, the resolution is limited and so is the frequency range of the PDS. Often the choices of bin sizes are arbitrarily made in the observer frame and lack considerations in astrophysics and instrument. Furthermore, the trend to use large bins (often from several to tens of microseconds) makes much information lost. Here we introduce a new kind of bin-free light curve, which is inspired here to avoid these information losses.
%From the above consideration, we determine to define a new light curve, which replaces the traditional representation of photon rate with reciprocal of time and avoids arbitrary prebinning. 
Actually, \citet{SCA08} has adopted a similar representation to detect quantum gravity photon dispersion in GRBs. Here we propose this method as an independent representation for the light curves for the following reasons. First, spectral time lag is not the subject in this paper. Instead, we focus on the detection of quasi-periodic signals in time profile through calculating its PDS. Second, we would like to generalize the application of the new light curve, e.g., identifying pulse which will be discussed in the discussion part. In fact, we believe the new light curve can substitute the old one to some extent and provides a different thread for analysis. Furthermore, this method is not supposed to be restricted within GRBs, and is suitable for various celestial objects including GRBs, X-ray binaries and pulsars, especially weak sources for the little information lost.

With some minor differences from Scargle's representation, we realize the method as follows. Assuming the interval between two neighboring photons has constant counting rate $1/dt_{i}$, where $dt_{i}$ is the width of this photon cell, we denote the time by averaging the successive arrival times, i.e., $(t_{i}+t_{i+1})/2$. As we do not have any interior information about the small cell, the simple assumption is temporarily reasonable for a rough estimate of the intensity and time. As demonstrated in Figure 1 by Scargle (2008), this cell-based representation inherits the salient features of variation while somewhat choppy. Selecting large cells ameliorates the choppiness, while leads to information losses. Here we do not care about the choppiness and tend to reserve as much information as possible, so small cell is preferred, which conduces to a PDS with large frequency range for later analysis. In some cases, several successive photons share the same recording time due to the limited time resolution of the instrument, which leads to a zero interval and the above formula nonsense. This problem is easily solved by increasing gradually the number of photons in the corresponding cell until it strides across the zero interval. So the more general equation of counting rate is
\begin{equation}
X_{n}=\frac{{\rm number\; of\; photons\; in\; cell}\; n}{{\rm size\; of\; cell}\; n}.
\end{equation}
It is noted that the numerator does not include the first photon in cell n, because it has been calculated in the last cell and is not counted as an increased one in cell n. The representation of time profile described above is defined as the new light curve and used throughout the rest of this paper. The upper-left panel of Figure 1 is an example of the new light curve with a set of simulated data.

To detect a periodic component in a noisy time series, the most standard tool is the Fourier transform and in most cases, the fast Fourier transform (FFT). However, the new light curve derived above is not evenly spaced, and FFT is not suitable. Instead we use Lomb periodogram to treat the unevenly spaced data \citep{LOM76,SCA82}. Futhermore, as in later applications we focus more on high-energy light curves whose photons are relatively poor, Lomb periodogram is very efficient for its good performance when the data size is relatively small so as to minimize aliasing (window border) distortions at the extremities of the time series. The Lomb periodogram analysis performs local least-squares fit of the data by sinusoids centered on each data point of the time series. Suppose that there are $N$ data points $X_{i}=X(T_{i}), i=1,...,N$. Then the mean and variance of the data are given by
\begin{equation}
\bar{X}\equiv\frac{1}{N}\sum_{i=1}^{N} X_{i}, 
\end{equation} and
\begin{equation}
 \sigma ^{2}\equiv\frac{1}{N-1}\sum_{i=1}^{N} (X_{i}-\bar{X} )^{2}.
\end{equation}
Then, the Lomb normalized periodogram (spectral power as a function of angular frequency $\omega \equiv 2\pi f> 0$) is defined by \citep{PRE94}
\begin{equation}
P_{N}(\omega )\equiv \frac{1}{2\sigma ^{2}} \left\{ \frac{[\sum_{i} (X_{i}-\bar{X})\cos{\omega (T_{i}-\tau )}] ^{2}}{\sum_{i} \cos^{2}{\omega (T_{i}-\tau )}}+\frac{[\sum_{i} (X_{i}-\bar{X}) \sin{\omega (T_{i}-\tau )})] ^{2}}{\sum_{i} \sin^{2}{\omega (T_{i}-\tau)}} \right\},
\label{eq:PDS}
\end{equation}
where $\tau $ is defined by
$\tan{2\omega \tau }=\frac{\sum_{i} \sin{2\omega T_{i}}}{\sum_{i} \cos{2\omega T_{i}}}$.
The normalized Lomb periodogram $P_{n}(\omega)$ is similar to an FFT power spectrum in which the presence of peaks at certain frequencies indicates the possible existence of periodic components. 

To assess the statistical significance of a given peak, the false-alarm probability of height $z$ is defined as 
\begin{equation}
Pr(>z)\approx Me^{-z},
\end{equation}
providing that the signal is a pure noise, namely the noise is independently normally distributed, and there exists $M$ independent frequencies. However, the ubiquitous non-Gaussian non-white noise invalidates the above assessment, such as in GRBs. Some more general results of Monte Carlo tests on the distribution of peaks in the Lomb periodogram for various types of processes are discussed in \citet{ZS01}.

\section{Application to the Data}

To verify the validity of our method, we perform some simulations. We start with a simple two-component light curve. The two periodic signals are both with the function form $A|\sin (\pi t/T)|$. They have periods of $T_{s}=10\pi s$ for the slow component and $T_{f}=\pi s$ for the fast component, and the amplitude ratio between them is $A_{s}:A_{f}=2:1$. Assuming the probability of detecting photons in the smallest time cell, i.e., the highest resolution of the instrument, proportional to the linear superposition of signal intensities, we get a time series data, where number one denotes a detection of photons and zero denotes no detection. For the nonsense of zeros in the data, we remove them and finally achieve a simplified TTE data which only consists of two periodic signals. The upper left and right panels in Figure 1 show the new light curve and binned version of the simulated time series, both of which clearly show the two components and confirm our creation of TTE data. The lower left and right panels show the PDS of TTE data with Lomb periodogram and binned light curve with FFT method, respectively. Two peaks that correspond to the two frequencies ($f_{s}=1/10\pi s^{-1}$ and $f_{f}=1/\pi s^{-1}$) are clearly identified in the left panel, while the slow component is not so clear in the right one, which might be influenced by the low-frequency slope. \citet{BEL98} have pointed out that the time binning suppresses the PDS by a factor related to the bin-size chosen \citep[cf.][]{VAN89}. Otherwise, The low-frequency power is somewhat unreliable for the distortions at the extremities of the time series, and especially vulnerable to different algorithms. Therefore, some more detailed analysis is needed to clarify the presence of the low-frequency slope. In the Lomb periodogram, the strong window effect, visible in the structure of sidelobes around both peaks, is due to the finite total interval over which the data is sampled. We also add some white noise to the mock light curve. We find that even when the amplitude of the white noise is comparable to the signal, two corresponding peaks still show up in our method. This suggests that our method is powerful in identifying the low frequency component.

We then apply our method to real observations including pulsar, X-ray binary and GRBs. 
%First, with evident period near 33ms, Crab Pulsar is selected to verify our method. Then in order to further confirm, we also try to find out QPOs in X-ray binary. Finally, we choose two GRBs: one is GRB 090709A, which has a known period 8.06s though is still in argument \citep{CEN10}; the other is GRB 090709B, with no periodicity found before \citep{MEE09}.
Since the relatively steady period in Crab Pulsar is well known, we choose a certain X-ray data from PCA (Proportional Counter Array) on-board RXTE (Rossi X-ray Timing Explorer) covering several minutes spanning from 23:36:33.4 UTC on 2011-12-31 until 23:49:27 UTC at the same day. In Figure 2, we can easily identify the peak near 30Hz, which is consistent with the known period. The other peaks located at integer multiples of fundamental frequency are simply the higher overtones of the signal.

As QPO is a very common phenomenon in X-ray binaries, we select a black-hole binary XTE J1550-564 with strong low-frequency oscillation. The archival data of XTE J1550-564 is from PCA instrument onboard the RXTE satellite on 1998-09-10. Since the QPO evolves with the state transition of source, we use simultaneously event-mode data and binned data to compare the result of our method with traditional FFT. From Figure 3a and 3b, the significant low-frequency QPO discovered by both methods agree with each other quite well while some harmonic peaks also appear here. The lower frequencies from the FFT method also shows a decreasing slope as in the lower-right panel of Figure 1, which might not be real signal but related to the binning process. Because the bin size of predefined light curve available is in microseconds, much larger than the time resolution of the instrument, the PDS of FFT gives a much narrower frequency range, which might miss some high-frequency signal.
%As the highest capable frequency is roughly $\frac{1}{\overline{\Delta t}}$ in our method, where $\overline{\Delta t}$ is the average time separation between successive photons (could be different for a larger $j$ choosing), and roughly reciprocal of the bin size in the FFT method, but the average separation is generally much less than the bin size, our method can achieve a higher frequency.

GRB 090709A was detected by the Burst Alert Telescope (BAT: \citet{BAR05}) on-board the \emph{Swift} satellite \citep{GEH04} at 7:38:34 UTC on 9 July 2009 \citep{MOR09}. The 15-350keV BAT light curve has been plotted by \citet{CEN10}, binned with 1s time resolution. A broad peak beginning at $t_{0}$ dominates the prompt emission of GRB 090709A and lasts about 100s. We select the period between $t_{0}-T_{90}$ and $t_{0}+2T_{90}$, where $T_{90}$ is taken from the second BAT catalogue \citep{SAK11}. In our analysis, we use the TTE data publicly available from the online database ftp://legacy.gsfc.nasa.gov/swift/data/. Figure 4a demonstrates the light curve obtained by the method described above. The time resolution in the light curve is so high that it looks very choppy. For some other tasks such as identifying pulses, large cells are preferred to moderate the choppiness. In traditional binned light curves, the background is usually not subtracted when calculating the PDS to keep the counts Poisson distributed and consequently the power distribution known, especially in the case of pure statistical noise, where the power is $\chi_{2}^{2}$ distributed. So we deal with the light curve in the same way. Figure 4b is the PDS by Lomb periodogram. In this figure, there indeed exists an ambiguous peak in the 0.1 Hz and 0.2 Hz with leakage to lower frequencies, which is consistent with the conclusion that this periodic signal is detected with only marginal significance ($\sim 2\sigma$) drawn by \citet{CEN10}. Compared with previous PDSs of GRBs, the power-law feature is still there (slope $\alpha \approx 2$), and overall trend of the curve is also similar: a power law or broken power law \citep{GUI12,DIC13}. These consistencies support again the reliability of our method.

As a contrast of GRB 090709A, we arbitrarily choose a short GRB 081223, detected by the Fermi Gamma-Ray Burst Monitor (GBM) at 10:03:57.148 UTC on 23 December 2008, to investigate the periodicity in short time series. For convenience of comparison, we also choose the duration between $t_{0}-T_{90}$ and $t_{0}+2T_{90}$ to calculate the light curve and PDS, which are respectively shown in Figure 5a and 5b. We can see from the upper panel that the prompt stage has higher and denser peaks. 
 We may use the peaks and clustering structure in the light curve to search for the pulse, which will be discussed in the discussion section. 
 %The step-like feather in the light curve comes from the highest time resolution, i.e., from the fits file, the time resolution is $10^{-4}$s, which is less probable to occur. If two photons come at the same time, the count rate goes to $10^4 {\rm s}^{-1}$, while if at the following recording time, there is no photon, the count rate goes to $1.5\times 10^{4} {\rm s^{-1}}$.
 The relatively smooth curve without any significant peak in the PDS plotted in Figure 5b agrees with former result \citep{DG13}. The PDS slopes are different between the two GRBs. Directly perceived from their PDSs, the one of GRB 090709A is shallower. This is consistent with the diversity of PDS slope \citep{BEL98}.
From Figures 1 and 3, considering the comparison of the Lomb periodogram and FFT, it seems the FFT always introduces an extra decreasing slope in the low-frequency portion of the PDS. That may indicate the overall slope (-5/3) of PDS in GRBs \citep{BEL98} is at least not all contributed by the data themselves.
Taking all the above into consideration, our method is justified to search for periodic features.

\section{Conclusion and Discussion}

We have developed a new representation of light curve and attempted to find out periodic features based on the new PDS by Lomb periodogram. Through application to the simulation of artificial light curve and the known real objects, we demonstrate that this method can identify significant periodic pulsations while reserve other features. It is no less powerful than the traditional binned light curve when seeking periodic components, as well as not limited by the bin size chosen. Especially in the case of weak sources like $\gamma$-ray light curve of pulsars, QPOs, GRBs, etc., where counts recorded are relatively poor in the short duration and variability time scales are considerably short, our method is more appropriate for its little information lost and much higher time resolution naturally produced, resulting in a broader frequency range in the PDS.

%After applying this method to several kinds of celestial objects as an effort to investigate periodicity, we have some findings summarized as follows:

%For evenly spaced data, there is a well-known natural set of frequencies, defined by
%$f_{n}=n/T,$
%where $n=-N_{0}/2,..., +N_{0}/2$, $T$ is the total time interval. And the highest frequency over which there is information is limited by the Nyquist critical frequency,$f_{N}=1/(2\Delta t),$
%where $\Delta t=T/N_{0}$ is the sampling interval, while $\Delta t$ is the shortest time interval. So the frequency range obtained from FFT is constrained by the bin size, which is usually exceedingly below the maximum resolution of instruments. However, for the case of uneven spacing, i.e. TTE data, $\Delta t$ is defined as $\Delta t=T/(N-1)=(t_{N}-t_{1})/(N-1),$
%which is much less than the bin size chosen. Therefore, as a valid and efficient tool, our method is capable of identifying a much wider frequency range than time-binned data, especially suitable for much higher frequencies. For those extremely poor photons recorded, or very high frequency objects, this method can not find the period. Folding might be a choice in this case.

In the realization of Lomb periodogram, there are two introduced pamameters adjustable. First, we specify how high in frequency to go, say $f_{hi}$. To choose $f_{hi}$, one way is to compare it with the Nyquist frequency $f_{c}$ which would obtain if the $N$ data points were evenly spaced over the same span $T$, that is $f_{c}=N/(2T)$. It is noted that a significant peak may be found (correctly) above the Nyquist frequency and without any significant aliasing down into the Nyquist interval, because the un-evenly spaced data has some points spaced much closer than the ``average" sampling rate, which removes ambiguity from any aliasing. That would not be possible for evenly spaced data. For the subject here, twice the Nyquist frequency is sufficient to go. The other parameter indicates the oversampling. In an FFT method, higher independent frequencies would be integer multiples of $1/T$. Since we are interested in the statistical significance of any peak that may occur in Lomb periodogram, however, it is better to (over-) sample more finely than at interval $1/T$, so that sample points lie close to the top of any peak. A value of four for the oversampling parameter is typically used. In spite of the differences from FFT described above, Lomb periodogram has been proved to be a good tool to use.

One of the most exciting application of our method is to detect quasi-periodicity in GRBs. It is widely believed that the central engine of GRBs is a rapidly rotating black hole with an hyper-accreting disk \citep{NP01,HJ03} or a millisecond highly magnetized neutron star \citep{UV92,TL13}. In both cases quasi-periodicity might exist, but so far no affirmative evidence has been detected from GRBs \citep{CEN10,DG13}. We want to search again for the periodicity with our method, especially in short GRBs, because Lomb periodogram is much less prone to aliasing distortions in short time series. Thus the marginal significance level of potential periodicity calculated before needs to be reconsidered and some new discovery is anticipated.      

Except for the PDS, the new light curve has some other considerable applications, like pulse identification, variability calculation, spectral time lag, etc. Considering the effect of noise, uncertainty of pulse shape plus arbitrariness of bin size, identifying the pulse in the binned light curve has always been a hard work, especially when the source is weak. While the unbinned light curve provides a new tool, the size of photon cells needs to be adjusted to reduce choppiness and a new criterion is also needed. In Figure 5a, the easily distinguished peaks and clustering structure may imply the existence of pulses. This suggests that the new light curve with smallest photon cells is also powerful when dealing with weak sources like short GRBs. 

Spectral time lag in different energy band in the light curve is also an interesting subject, which might be used to test the radiation mechanism, quantum gravity theories, medium on the line of sight, etc. In the cross-correlation theorem \citep{JW68}, the cross-correlation function is the inverse Fourier transform of the cross-spectrum. With this relation, spectral time lag can be calculated without the limitation of samplings, uniform bins and identical time span, and with high time resolution \citep{SCA89}. The energy bin may also be released together with the time bins in the spectral time lag searching.
%Photon dispersion is predicted by quantum gravity theories due to a possible dependence of the speed of light on photon energy. It has been searched in binned light curves in different energy bands using a cross-correlation methodology \citep{NOR02}. However, we cannot directly use this method because of the uneven time spacing in unbinned light curves.

%\acknowledgements
We thank Jianeng Zhou and Xiaofeng Cao for the instruction on data analysis. This work was supported by the National Natural Science Foundation of China (Grant No. U1231101), the National Basic Research Program (973 Program) of China (Grant No. 2014CB845800) and the Chinese-Israeli Joint Research Project (Grant No. 11361140349).

%\appendix
\newcommand\aj{{AJ}}%
          % Astronomical Journal
\newcommand\actaa{{Acta Astron.}}%
  % Acta Astronomica
\newcommand\araa{{ARA\&A}}%
          % Annual Review of Astron and Astrophys
\newcommand\apj{{ApJ}}%
          % Astrophysical Journal
\newcommand\apjl{{ApJ}}%
          % Astrophysical Journal, Letters
\newcommand\apjs{{ApJS}}%
          % Astrophysical Journal, Supplement
\newcommand\ao{{Appl.~Opt.}}%
          % Applied Optics
\newcommand\apss{{Ap\&SS}}%
          % Astrophysics and Space Science
\newcommand\aap{{A\&A}}%
          % Astronomy and Astrophysics
\newcommand\aapr{{A\&A~Rev.}}%
          % Astronomy and Astrophysics Reviews
\newcommand\aaps{{A\&AS}}%
          % Astronomy and Astrophysics, Supplement
\newcommand\azh{{AZh}}%
          % Astronomicheskii Zhurnal
\newcommand\baas{{BAAS}}%
          % Bulletin of the AAS
\newcommand\caa{{Chinese Astron. Astrophys.}}%
  % Chinese Astronomy and Astrophysics
\newcommand\cjaa{{Chinese J. Astron. Astrophys.}}%
  % Chinese Journal of Astronomy and Astrophysics
\newcommand\icarus{{Icarus}}%
  % Icarus
\newcommand\jcap{{J. Cosmology Astropart. Phys.}}%
  % Journal of Cosmology and Astroparticle Physics
\newcommand\jrasc{{JRASC}}%
          % Journal of the RAS of Canada
\newcommand\memras{{MmRAS}}%
          % Memoirs of the RAS
\newcommand\mnras{{MNRAS}}%
          % Monthly Notices of the RAS
\newcommand\na{{New A}}%
  % New Astronomy
\newcommand\nar{{New A Rev.}}%
  % New Astronomy Review
\newcommand\pra{{Phys.~Rev.~A}}%
          % Physical Review A: General Physics
\newcommand\prb{{Phys.~Rev.~B}}%
          % Physical Review B: Solid State
\newcommand\prc{{Phys.~Rev.~C}}%
          % Physical Review C
\newcommand\prd{{Phys.~Rev.~D}}%
          % Physical Review D
\newcommand\pre{{Phys.~Rev.~E}}%
          % Physical Review E
\newcommand\prl{{Phys.~Rev.~Lett.}}%
          % Physical Review Letters
\newcommand\pasa{{PASA}}%
  % Publications of the Astron. Soc. of Australia
\newcommand\pasp{{PASP}}%
          % Publications of the ASP
\newcommand\pasj{{PASJ}}%
          % Publications of the ASJ
\newcommand\qjras{{QJRAS}}%
          % Quarterly Journal of the RAS
\newcommand\rmxaa{{Rev. Mexicana Astron. Astrofis.}}%
  % Revista Mexicana de Astronomia y Astrofisica
\newcommand\skytel{{S\&T}}%
          % Sky and Telescope
\newcommand\solphys{{Sol.~Phys.}}%
          % Solar Physics
\newcommand\sovast{{Soviet~Ast.}}%
          % Soviet Astronomy
\newcommand\ssr{{Space~Sci.~Rev.}}%
          % Space Science Reviews
\newcommand\zap{{ZAp}}%
          % Zeitschrift fuer Astrophysik
\newcommand\nat{{Nature}}%
          % Nature
\newcommand\iaucirc{{IAU~Circ.}}%
          % IAU Cirulars
\newcommand\aplett{{Astrophys.~Lett.}}%
          % Astrophysics Letters and Communications
\newcommand\apspr{{Astrophys.~Space~Phys.~Res.}}%
          % Astrophysics Space Physics Research
\newcommand\bain{{Bull.~Astron.~Inst.~Netherlands}}%
          % Bulletin Astronomical Institute of the Netherlands
\newcommand\fcp{{Fund.~Cosmic~Phys.}}%
          % Fundamental Cosmic Physics
\newcommand\gca{{Geochim.~Cosmochim.~Acta}}%
          % Geochimica Cosmochimica Acta
\newcommand\grl{{Geophys.~Res.~Lett.}}%
          % Geophysics Research Letters
\newcommand\jcp{{J.~Chem.~Phys.}}%
          % Journal of Chemical Physics
\newcommand\jgr{{J.~Geophys.~Res.}}%
          % Journal of Geophysical Research
\newcommand\jqsrt{{J.~Quant.~Spec.~Radiat.~Transf.}}%
          % Journal of Quantitiative Spectroscopy and Radiative Trasfer
\newcommand\memsai{{Mem.~Soc.~Astron.~Italiana}}%
          % Mem. Societa Astronomica Italiana
\newcommand\nphysa{{Nucl.~Phys.~A}}%
          % Nuclear Physics A
\newcommand\physrep{{Phys.~Rep.}}%
          % Physics Reports
\newcommand\physscr{{Phys.~Scr}}%
          % Physica Scripta
\newcommand\planss{{Planet.~Space~Sci.}}%
          % Planetary Space Science
\newcommand\procspie{{Proc.~SPIE}}%
          % Proceedings of the SPIE
\let\astap=\aap
\let\apjlett=\apjl
\let\apjsupp=\apjs

\begin{figure}
\includegraphics[angle=0,scale=.25]{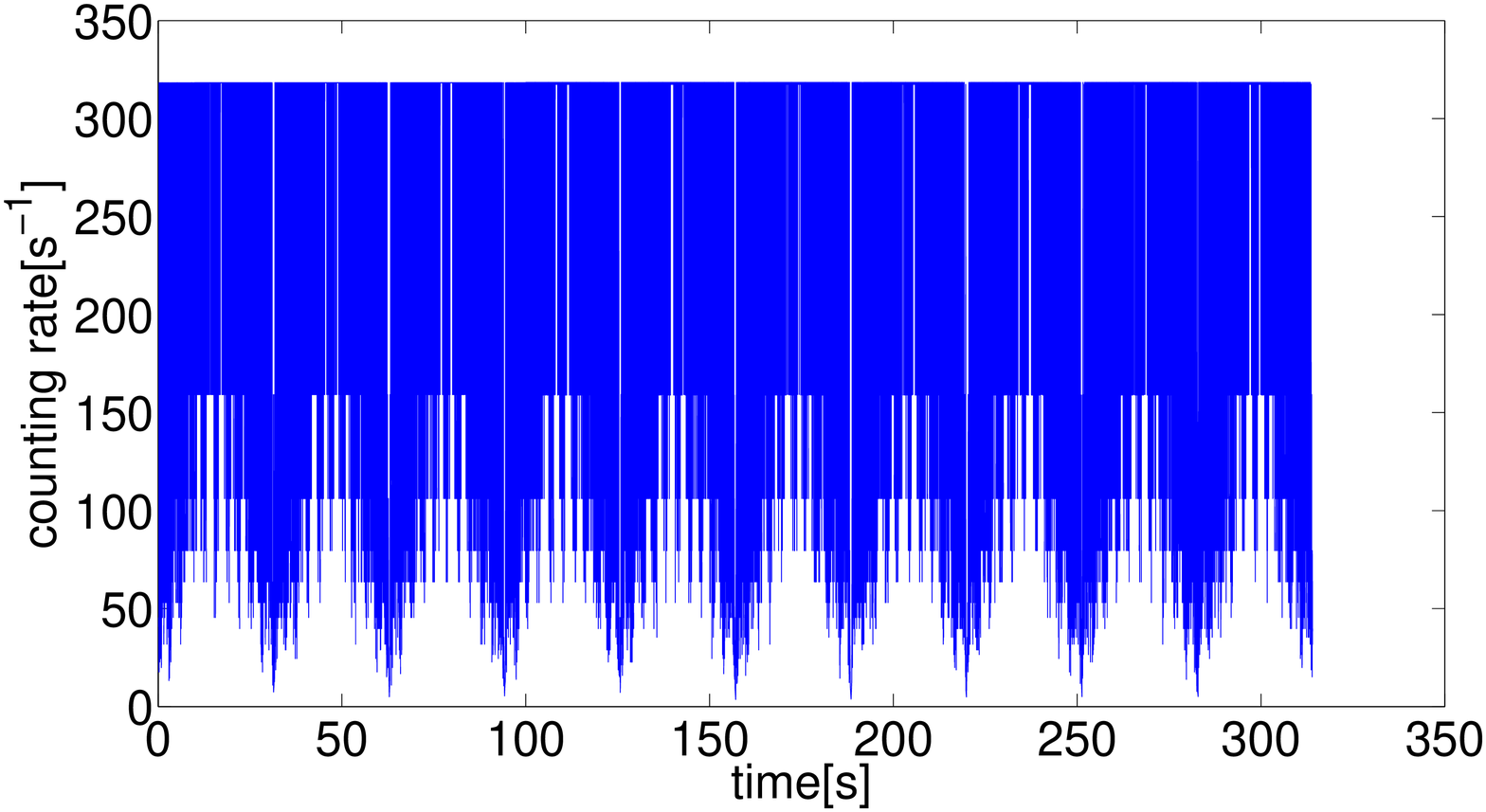}
\includegraphics[angle=0,scale=.25]{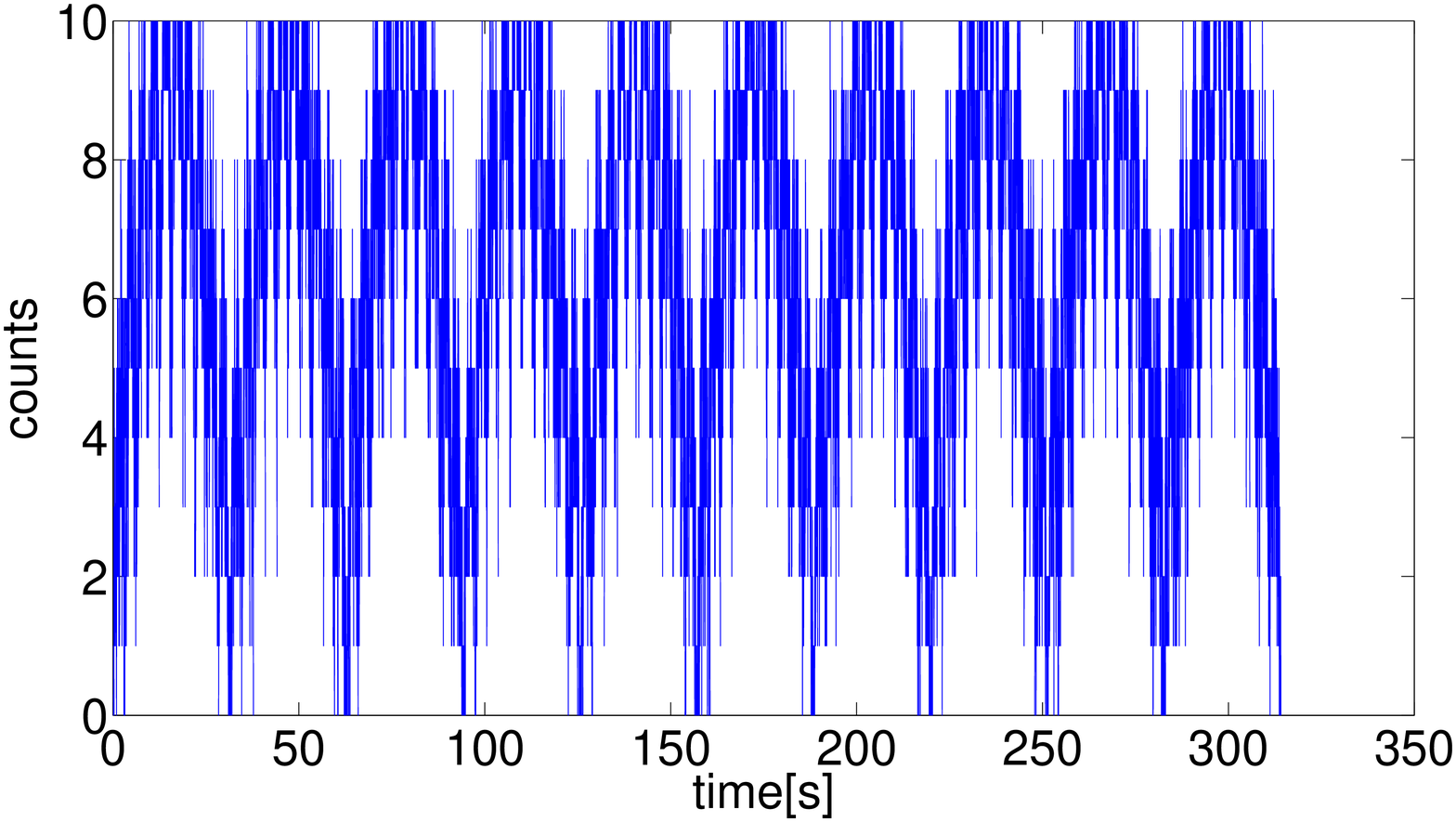}
\includegraphics[angle=0,scale=.25]{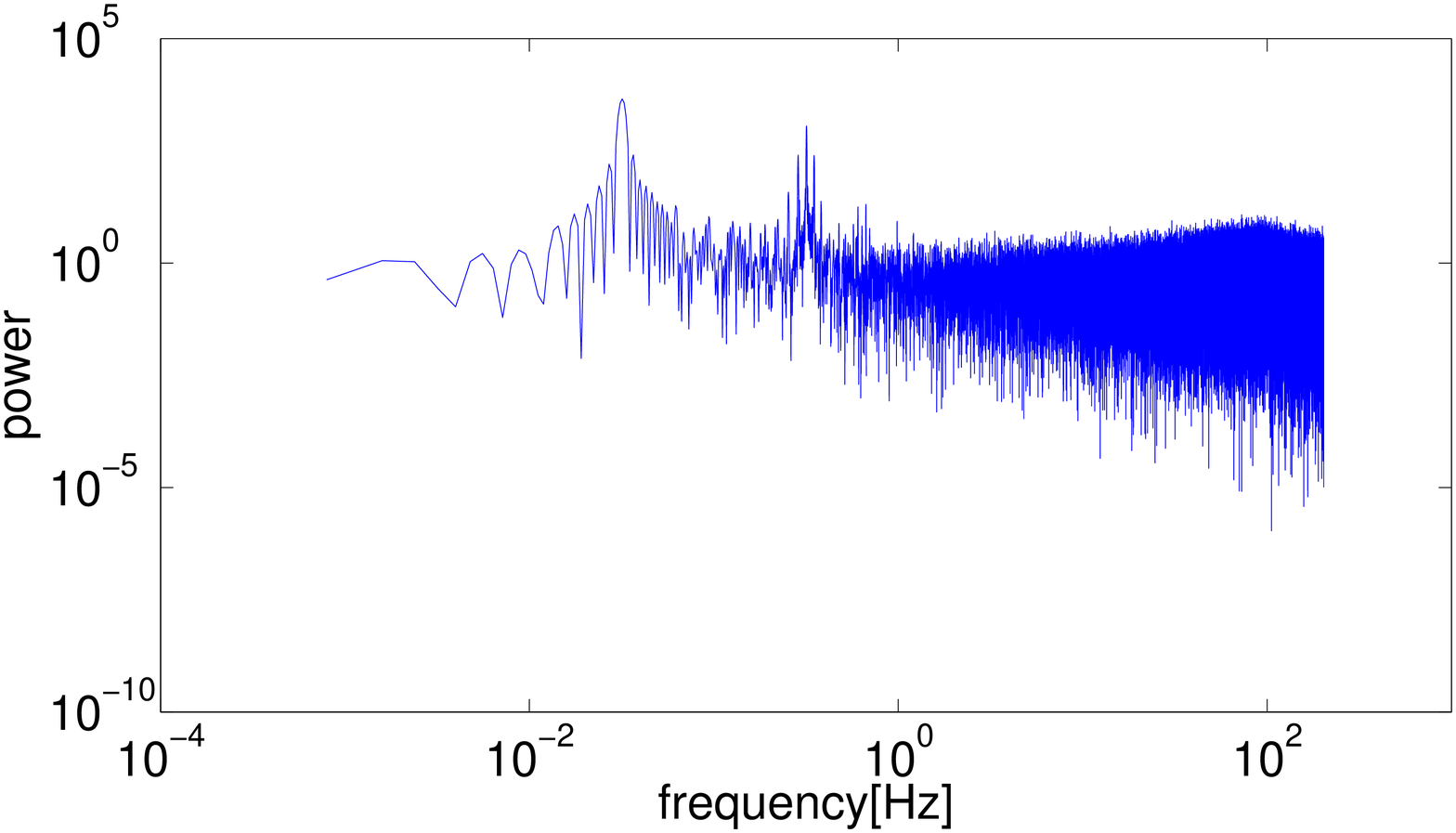}
\includegraphics[angle=0,scale=.25]{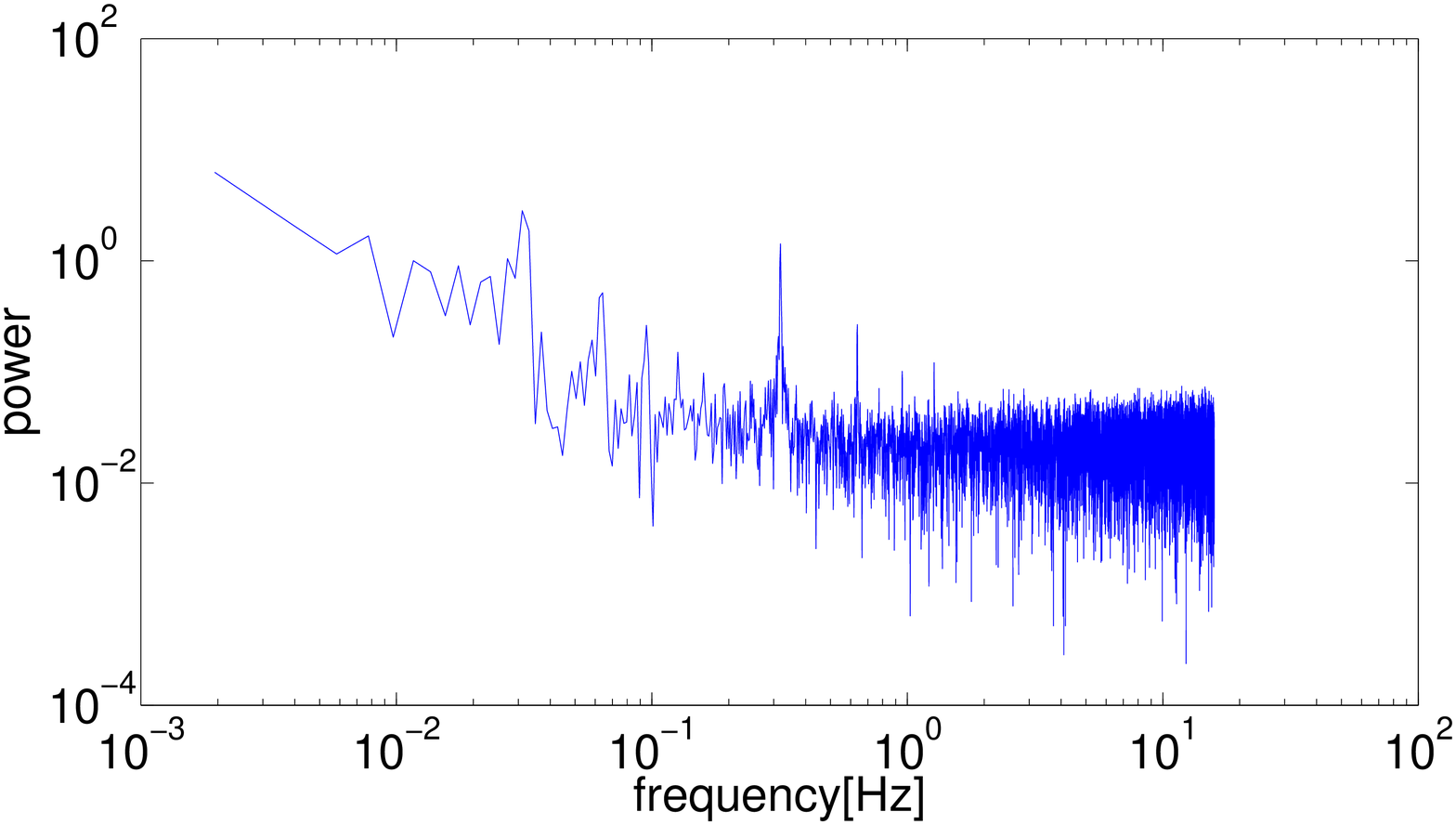}
\caption{The simulation test. \textit{Top panel:} The new light curve (left) and binned light curve (right); \textit{Lower panel:} The PDS by Lomb periodogram (left) and FFT (right). The simulated signal spans ten times the period of the slow component, i.e. 100$\pi$s, with a time resolution of $\pi/1000$s. The bin size in the binned light curve is ten times the time resolution, i.e. $\pi/100$s.}
\end{figure}

\clearpage

\begin{figure}
\includegraphics[angle=0,scale=.35]{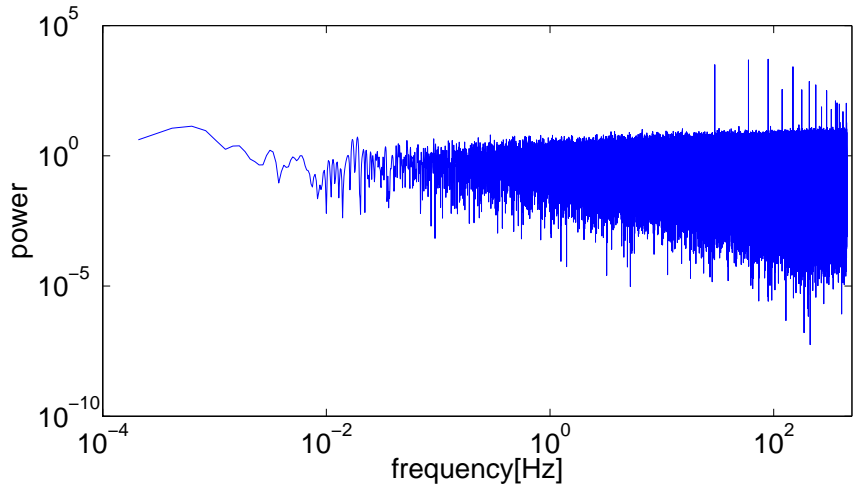}
\caption{PDS of Crab Pulsar by Lomb periodogram.}
\end{figure}

\begin{figure}
\includegraphics[angle=0,scale=.35]{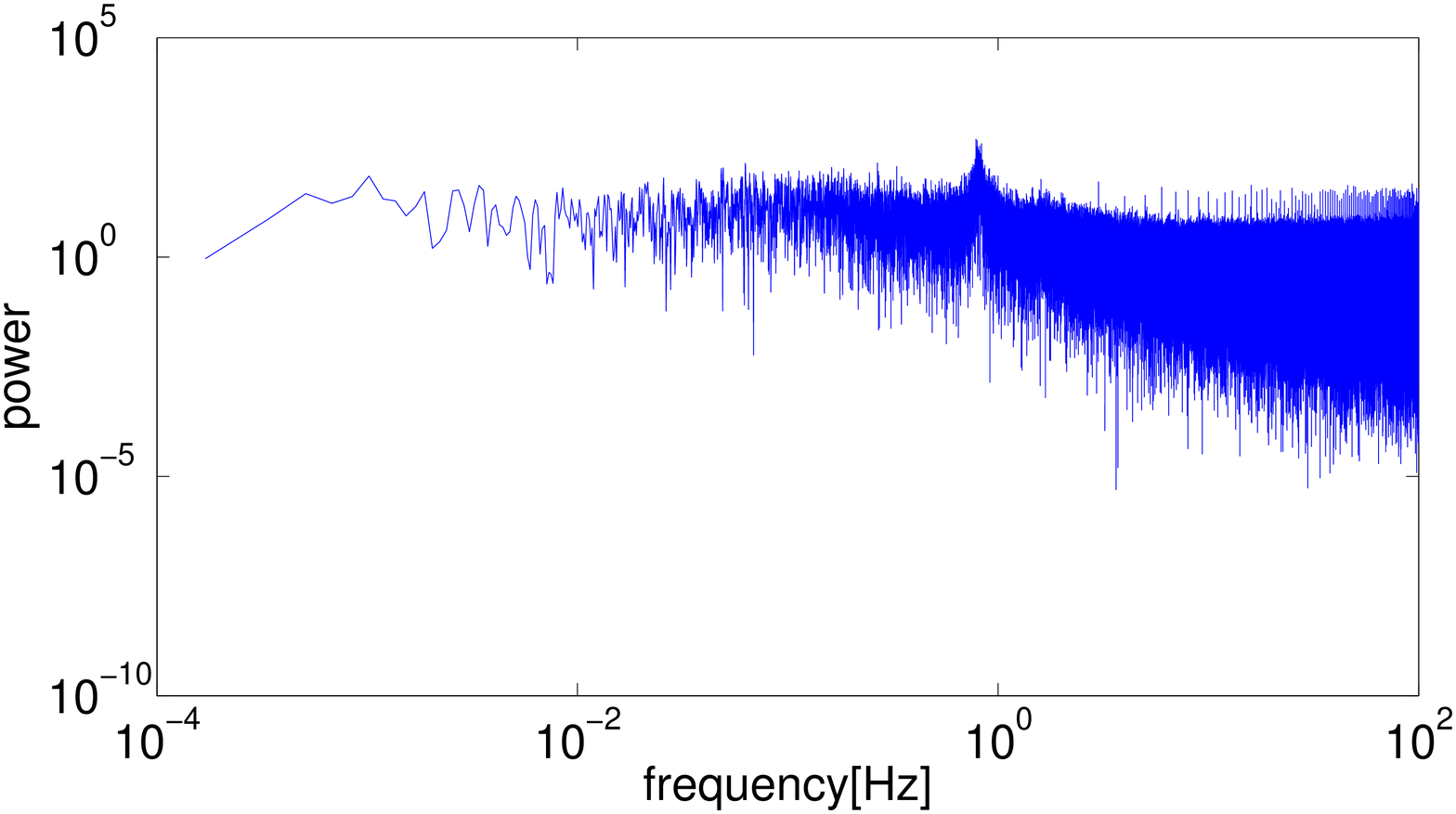}
\includegraphics[angle=0,scale=.35]{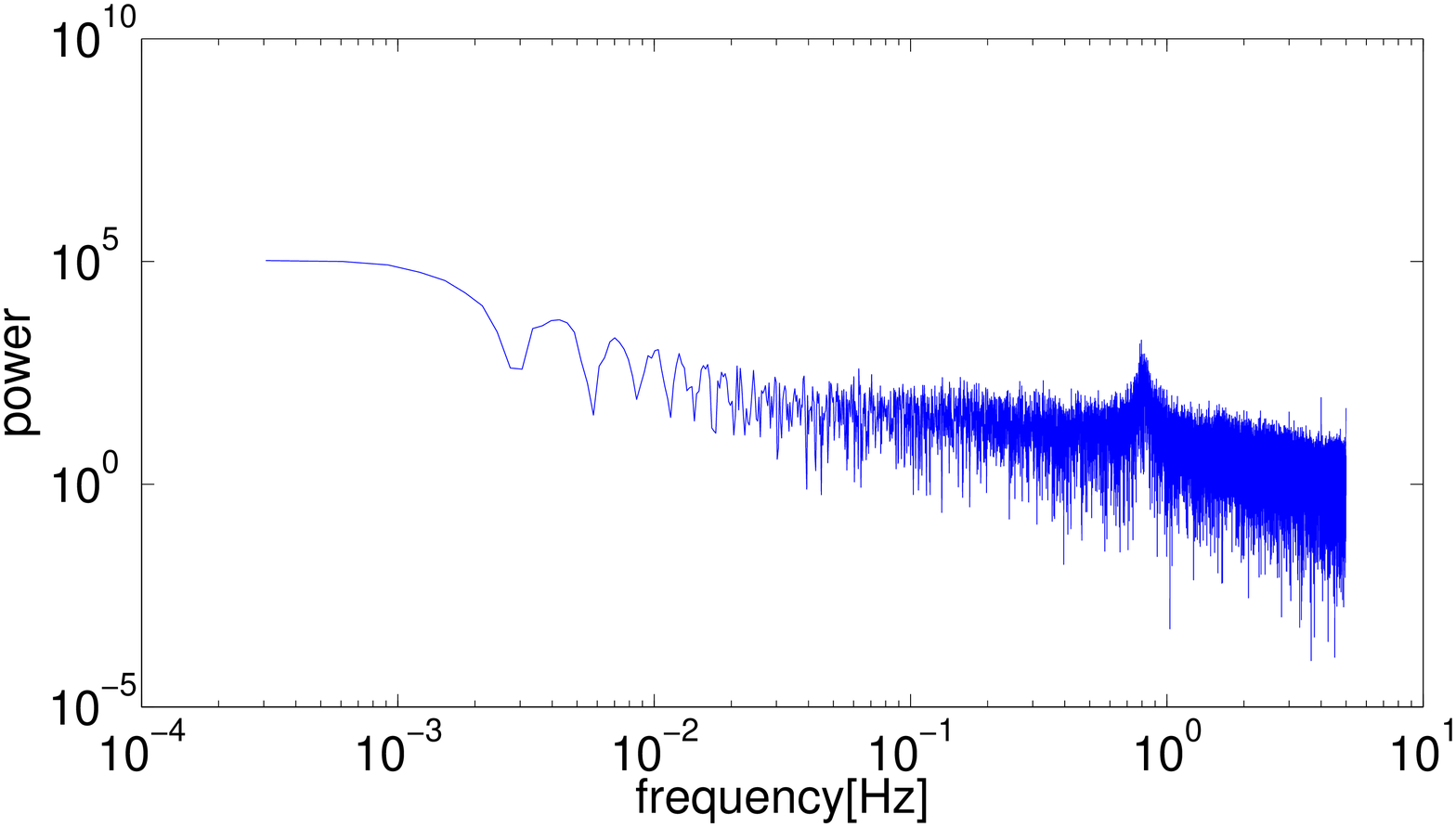}
\caption{\textit{top panel:} PDS of XTE J1550-564 obtained by using unbinned light curve. \textit{lower panel:} PDS of the same source using binned light curve.}
\end{figure}

\begin{figure}
\includegraphics[angle=0,scale=.35]{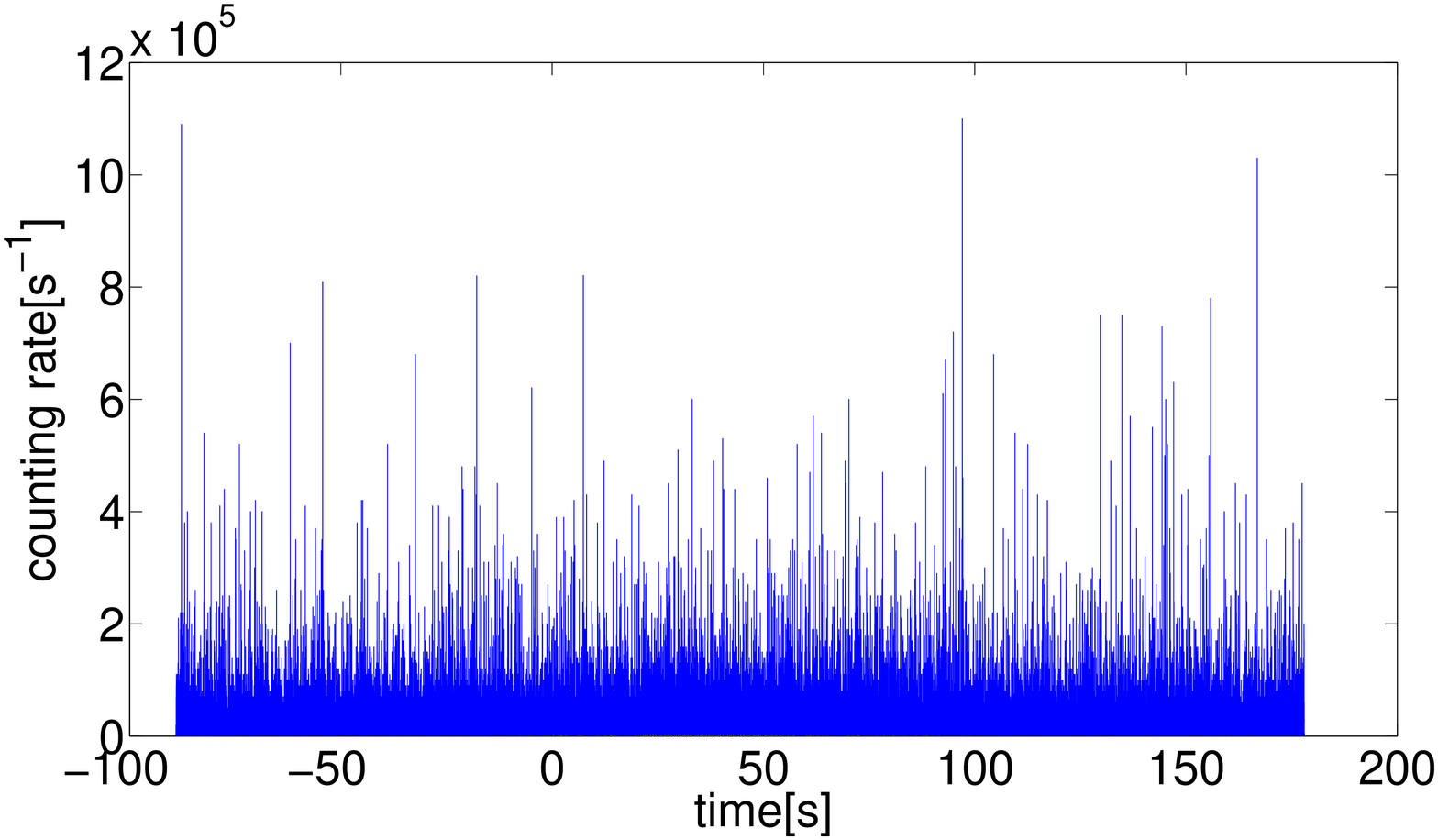}
\includegraphics[angle=0,scale=.35]{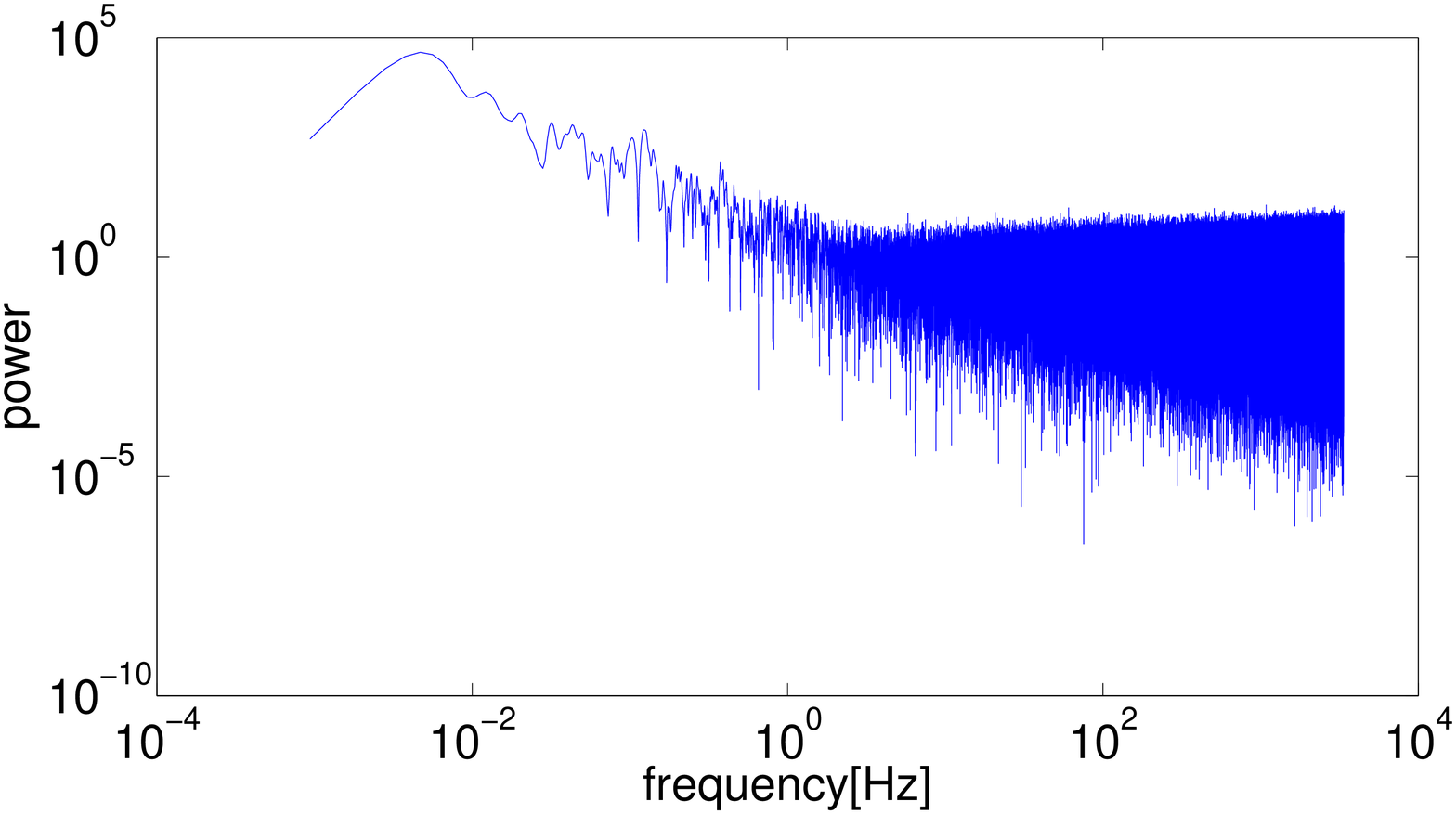}
\caption{\textit{top panel:} unbinned light curve of GRB 090709A. \textit{lower panel:} PDS of the unbinned light curve.}
\end{figure}

\begin{figure}
\includegraphics[angle=0,scale=.35]{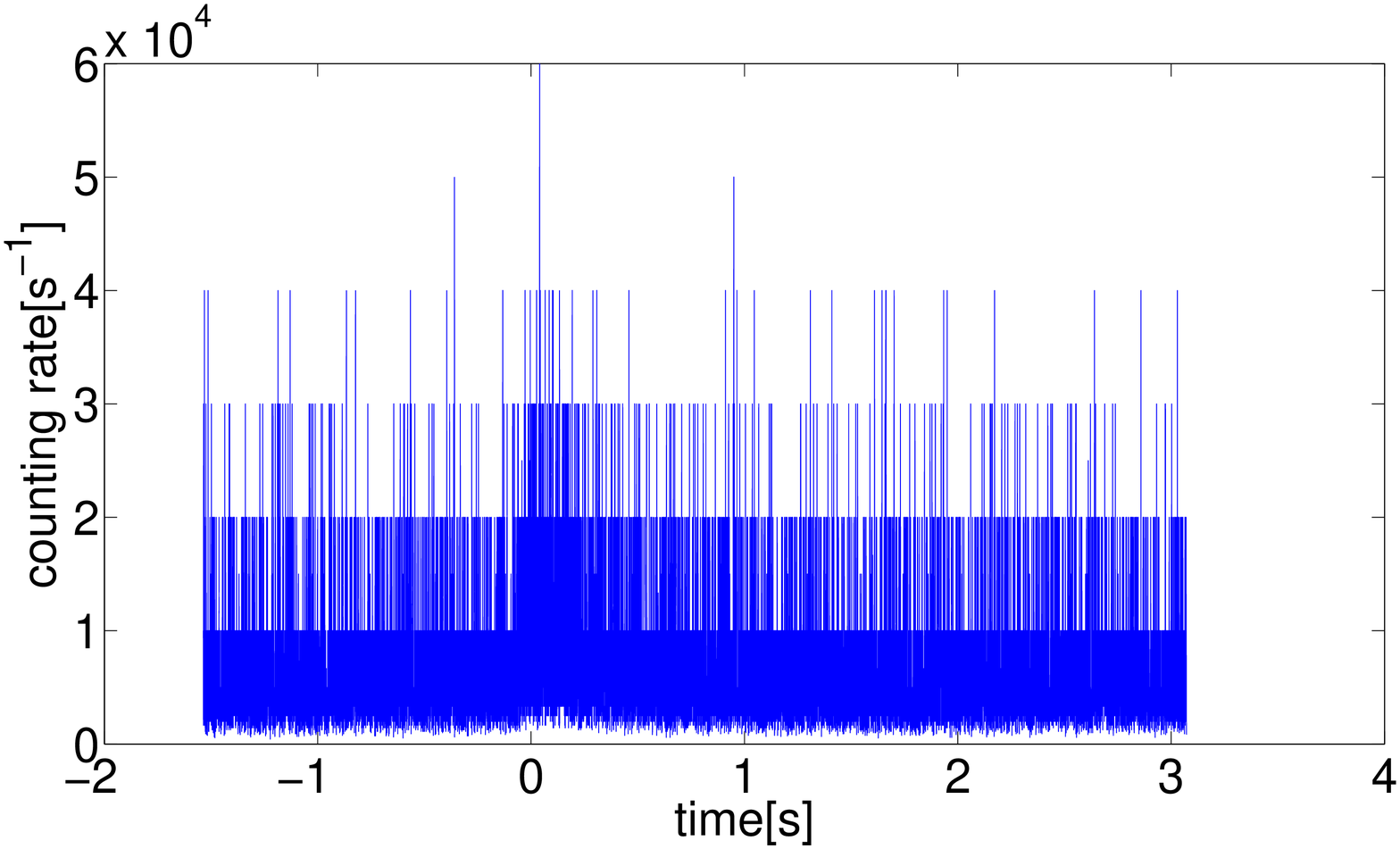}
\includegraphics[angle=0,scale=.35]{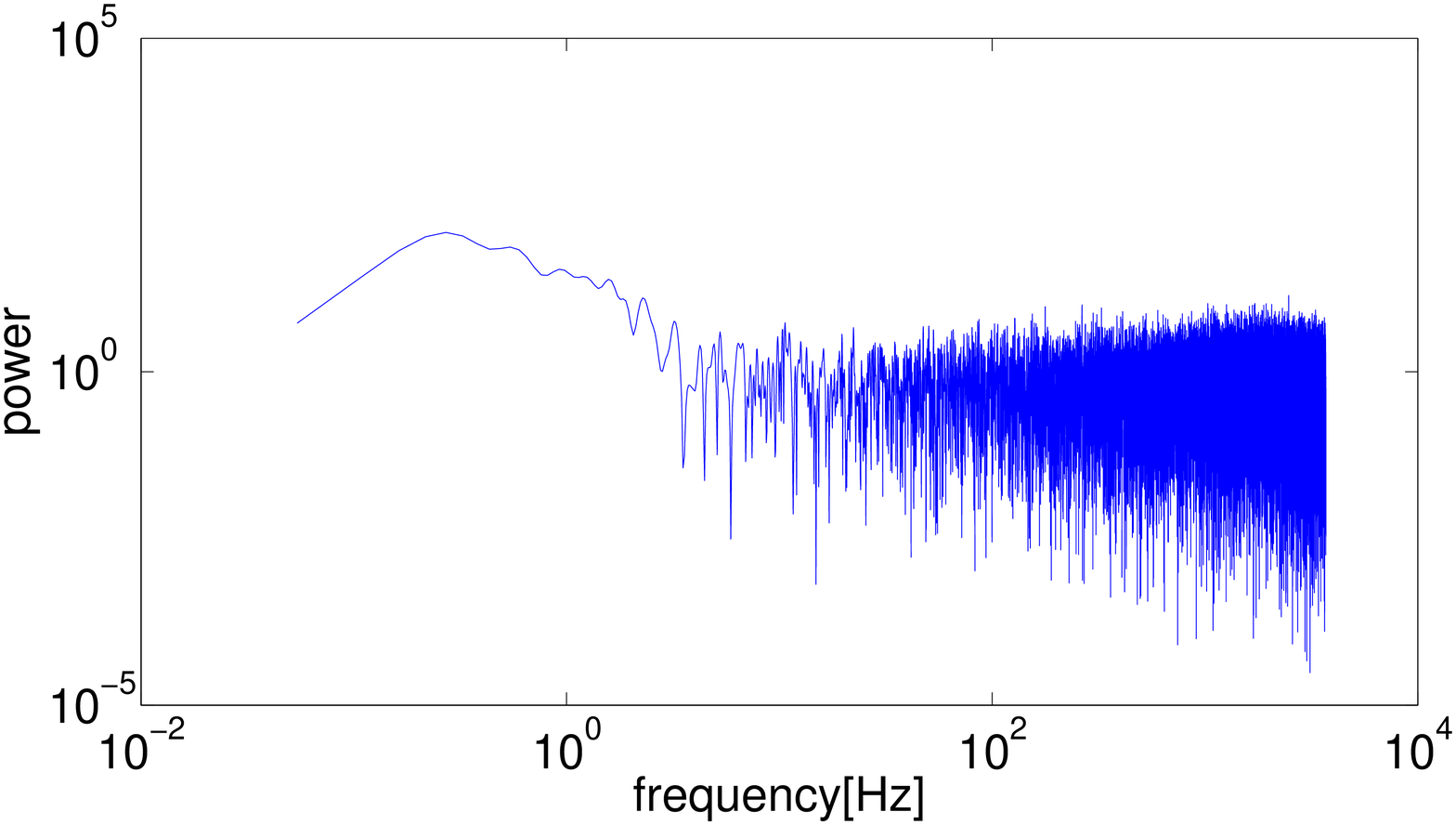}
\caption{\textit{top panel:} unbinned light curve of GRB 081223. \textit{lower panel:} PDS of the light curve.}
\end{figure}

\end{document}